\documentclass[aps,pra,showpacs,floatfix,superscriptaddress,twocolumn]{revtex4}
\usepackage[dvips]{graphicx}
\usepackage{color}

\usepackage{longtable}
\usepackage{dcolumn}
\usepackage[dvips]{graphicx}
\usepackage{bm}
\usepackage{bbm}

\usepackage{times}
\usepackage{nicefrac}
\usepackage{amsmath}
\usepackage{amsfonts}
\usepackage{amssymb}
\usepackage{amsthm}

\newcolumntype{.}{D{x}{}{-1}}

\newcommand{\vare}{\varepsilon}

\newcommand{\lbr}{\langle}
\newcommand{\rbr}{\rangle}

\begin{document}

\title{Relativistic
configuration-interaction calculation of $\bm{K\alpha}$ transition energies in beryllium-like
argon}

\author{V. A. Yerokhin}

\affiliation{Center for Advanced Studies, St.~Petersburg State Polytechnical University,
 195251 St.~Petersburg, Russia}

\affiliation{Helmholtz-Institut Jena, D-07743 Jena, Germany}

\author{A. Surzhykov}

\affiliation{Helmholtz-Institut Jena, D-07743 Jena, Germany}

\author{S. Fritzsche}

\affiliation{Helmholtz-Institut Jena, D-07743 Jena, Germany}
\affiliation{{Theoretisch-Physikalisches Institut, Friedrich-Schiller-Universit\"at Jena, D-07743
Jena, Germany}}

\begin{abstract}

Relativistic configuration-interaction calculations have been performed for the energy levels of
the low-lying and core-excited states of beryllium-like argon, Ar$^{14+}$. These calculations
include the one-loop QED effects as obtained by two different methods, the screening-potential
approach as well as the model QED operator approach. The calculations are supplemented by a
systematic estimation of uncertainties of theoretical predictions.

\end{abstract}


\maketitle

\section{Introduction}

The subject of the present investigation are the ${K\alpha}$ transitions in beryllium-like ions. In
these transitions, an electron changes the principal quantum number from $n = 2$ to $n = 1$,
filling the vacancy in the $K$-shell core in the initial state. These transitions contribute to the
most prominent $K$-shell emission lines of highly charged ions which have been detected in the
spectra of nearly all classes of cosmic X-ray sources. In laboratories, moreover, the $K$-shell
emission lines are often used for the diagnostics of hot plasma, notably in magnetic nuclear fusion
and tokamaks, and thus help getting useful information about the equilibrium and non-equilibrium
charge-state distributions of ions as well as about the electron and ion temperatures. In view of
this importance of the $K\alpha$ line for astrophysics and laboratory diagnostics, accurate
theoretical predictions are needed for the reliable identification and interpretation of
experimental spectral data \cite{beiersdorfer:93,wargelin:01}.

In our previous investigations, we performed relativistic configuration-interaction calculations of
the $K\alpha$ transitions in lithium-like ions \cite{yerokhin:12:lilike} and in beryllium-like iron
\cite{yerokhin:14:belike}. In the current work, we extend our calculations to beryllium-like argon
to accommodate ongoing experiments on this ion \cite{amaro:12,szabo:13,indelicato:14}.

The paper is organized as follows. In the next section, we give a brief outline of our computation
method. Sec.~\ref{sec:res} presents the results of our calculations and compares them with the
previous theoretical and experimental data. Relativistic units $\hbar=c=1$ and charge units
$e^2/4\pi  = \alpha$ are used throughout this paper.

\section{Method of calculation}

Within the relativistic quantum mechanics, the energy of the system can be determined by solving
the secular equation
\begin{eqnarray}\label{eq:000}
    {\rm det} \bigl\{\lbr \gamma_r PJM|H_{\rm DCB}|\gamma_s PJM\rbr -E_r\,\delta_{rs}\bigr\} =
    0\,,
\end{eqnarray}
where ``det'' denotes the determinant of the matrix. $H_{\rm DCB}$ is the no-pair
Dirac-Coulomb-Breit (DCB) Hamiltonian ,
\begin{eqnarray}
    H_{\rm DCB} = \sum_i h_{\rm D}(i) + \sum_{i<j} \left[ V_{C}(i,j)+
    V_{B}(i,j)\right]\,,
\end{eqnarray}
where the indices $i,j = 1,\ldots,N$ numerate the electrons, $h_D$ is the one-particle
Dirac-Coulomb Hamiltonian, and $V_{C}$ and $V_B$ are the Coulomb and the Breit parts of the
electron-electron interaction. It is assumed that $H_{\rm DCB}$ acts in the space of the wave
functions constructed from the positive-energy eigenfunctions of some one-particle Dirac
Hamiltonian (the so-called no-pair approximation).

The $N$-electron wave function $\Psi \equiv \Psi(PJM)$ is assumed to have a definite parity $P$,
total angular momentum $J$, and angular momentum projection $M$. In the CI method, the
eigenfunctions $\Psi(PJM)$ are represented by a finite sum of the configuration-state functions
(CSFs) with the same $P$, $J$, and $M$,
\begin{equation}\label{eq4}
  \Psi(PJM) = \sum_r c_r \Phi(\gamma_r PJM)\,,
\end{equation}
where $\gamma_r$ denotes the set of additional quantum numbers that determine the CSF. The CSFs are
constructed as linear combinations of antisymmetrized products of one-electron orbitals $\psi_n$.
The linear coefficients $c_r$ in Eq.~(\ref{eq4}) and the energy of the corresponding atomic state
are obtained by solving the secular equation (\ref{eq:000}).

The elements of the Hamiltonian matrix are typically calculated as linear combinations of one- and
two-particle radial integrals,
\begin{align}\label{eq7}
\lbr \gamma_r PJM| &\,  H_{\rm DCB}|\gamma_s PJM\rbr = \sum_{ab}
 d_{rs}(ab)\,I(ab)
  \nonumber \\
 + &\, \alpha \sum_k \sum_{abcd} v_{rs}^{(k)}(abcd)\,
  R_{k}(abcd)\,.
\end{align}
Here, $a$, $b$, $c$, and $d$ numerate the one-electron orbitals, $d_{rs}$ and $v^{(k)}_{rs}$ are
the angular coefficients, $I(ab)$ are the one-electron radial integrals, and $R_{k}(abcd)$ are the
two-electron radial integrals. Details of our implementation of the CI method  are described in the
previous papers \cite{yerokhin:12:lilike,yerokhin:14:belike}.

In order to obtain accurate theoretical predictions for energy levels, the relativistic energies
obtained from the DCB Hamiltonian should be supplemented by the QED corrections. The best accuracy
can be presently obtained within the {\em ab initio} QED method \cite{shabaev:02:rep}. Practical
calculations within this approach started in late 1990s \cite{artemyev:97,yerokhin:97:pla}. Because
of large technical difficulties, however, such {\em ab inito} QED calculations are still restricted
mainly to low-lying states of helium-like and lithium-like ions \cite{artemyev:05:pra,kozhedub:10}.
For calculations of more complex atomic systems, one has to rely on one of the simplified
treatments of QED effects.

In the present work, we describe the QED effects by means of two different approximate methods. By
comparing the results from these approaches, we estimate the uncertainty of our treatment. The
first method is based on summing up the self-energy and vacuum-polarization QED corrections
calculated for each one-electron orbital in an effective screening potential. The total QED
correction for a given many-electron state is then obtained by adding the QED contributions from
all one-electron orbitals, weighted by their fractional occupation numbers as obtained from the
eigenvectors of the CI calculation. In this method, the QED correction $\delta E_{\rm QED}$ is
given by
\begin{align}
\delta E_{\rm QED} = \sum_a q_a\,\bigl[\lbr a| \Sigma_{\rm SE}(\vare_a)|a\rbr
+  \lbr a| V_{\rm VP} |a\rbr  \bigr]\,,
\end{align}
where the index $a$ runs over all one-electron orbitals contributing to the many-electron state of
interest, $q_a$ is the occupation number of the one-electron orbital, $\Sigma_{\rm SE}$ is the
self-energy operator, $\vare_a$ is the Dirac energy of the one-electron state $a$, and $V_{\rm VP}$
is the vacuum polarization potential. The numerical method for calculating the one-loop self-energy
and vacuum-polarization matrix elements in a general screening potential used in the present work
was developed in Ref.~\cite{yerokhin:11:fns}. This approach for treatment of the Lamb shift in
atoms was used in our previous studies \cite{yerokhin:12:lilike,yerokhin:14:belike} and similarly
by several other authors, and for beryllium-like ions in particular by Chen and Cheng
\cite{chen:97}. In the present work we use in fact two different screening potentials in which the
one-loop QED corrections are calculated, the core-Hartree (CH) potential and the localized
Dirac-Fock (LDF) potential. The definition of these potentials can be found in our previous works
\cite{yerokhin:12:lilike,yerokhin:08:pra}.

The second method for evaluation of the QED effects is based on the model QED operator $h^{\rm
QED}$ formulated recently by Shabaev et al.~\cite{shabaev:13:qedmod} and implemented in the QEDMOD
Fortran package \cite{shabaev:14:qedmod}. To this end, we added the model QED operator to the DCB
Hamiltonian by modifying the one-electron integrals $I(a,b)$ of Eq.~(\ref{eq7})  in our CI code by
\begin{align}
I(ab)\to I(ab) + \delta_{\kappa_a,\kappa_b}\,\lbr a|h^{\rm QED}|b\rbr\,,
\end{align}
and where $\kappa_a$ denotes the relativistic angular quantum number of the state $a$. If either
$a$ or $b$ is a continuum state (i.e., $\max(\vare_a,\vare_b) > m$), the matrix element of $h^{\rm
QED}$ is assumed to be zero. The QED correction to the energy level is then identified by taking
the difference of the CI eigenvalues with and without the $h^{\rm QED}$ operator.

Our calculations show that QED corrections obtained by the different methods are in good agreement
with each other. In the case of core-excited states, the difference between the results remains
well within the 1\% range. For the ground and valence-excited states, the deviation is slightly
larger, on the level of 1-2\%. This is explained by the relatively large effect of the mutual
screening of the $1s$ electrons by each other, which is not very well described by approximate
methods.

In the present work, we use the QED results obtained with the LDF potential as final values of the
QED correction and estimate its uncertainty by the maximal difference between the three QED values.

\section{Results and discussion}
\label{sec:res}

In Table~\ref{tab:main} we present the calculated energy levels of the beryllium-like argon,
Ar$^{14+}$. The total energy is given for the ground $1s^22s^2\,^1S$ state, whereas the {\em
relative} energies (with respect to the ground state) are given for the excited states. Our results
are compared with the NIST compilation based on experimental and theoretical data
\cite{nist:13,shirai:00}, with the relativistic many-body perturbation theory calculation by
Safronova {\em et al.} \cite{safronova:96} and with the multiconfigurational Dirac-Fock calculation
of Cota {\em et al.} \cite{costa:01}, as well as with the experimental results \cite{edlen:83}.

All our theoretical predictions are supplied with the uncertainties, which include the error
estimates for the Dirac-Coulomb-Breit energy as well as for the QED correction. The former estimate
is evaluated by analyzing the configuration-interaction results for successively enlarged basis
sets. The QED error was obtained by comparing 3 different QED values from different approaches. The
estimated accuracy of our theoretical energies of the core-excited states ranges from $10^{-5}$ to
$10^{-4}$ in relative units, or from 0.002 to 0.02~Rydbergs.

For the valence-excited states, our results are in good agreement with previous calculations
\cite{safronova:96}, experimental data \cite{edlen:83}, and with the NIST data base \cite{nist:13}.
For the core-excited states, however, there are only 3 entries available in the NIST data base and
apparently no experimental results. The only detailed theoretical study of these levels was
performed in Ref.~\cite{costa:01} by the multiconfigurational Dirac-Fock (MCDF) method. Our results
are in general agreement with those of Ref.~\cite{costa:01}, but the differences are well outside
our error bars. The largest deviations of about 0.1~Ry are found for for the $^5P_J$ states and for
the highest core-excited levels. For the other states, the differences are smaller, typically
within 0.05~Ry. It might be mentioned that our result for the $^1P^o_1$ state is in excellent
agreement with the preliminary result of the Paris experiment
\cite{amaro:12,szabo:13,indelicato:14}.

In Table~\ref{tab:trans} we present our theoretical results for the wavelength of the $K\alpha$
transition lines in beryllium-like argon.

\section{Conclusion}

In summary, we performed relativistic configuration-interaction calculations of the energy levels
of the ground, valence-excited and core-excited states in beryllium-like argon, Ar$^{14+}$. The
relativistic Dirac-Coulomb-Breit energies obtained by the configuration-interaction method were
supplemented with the QED energy shifts calculated separately. The QED corrections were obtained by
two different approximate methods, the screening-potential approach and the model QED operator
method. From the comparison of the results of these two approaches we estimated the uncertainty of
the overall QED shift. The uncertainty of the Dirac-Coulomb-Breit energies was evaluated by
analysing the convergence of the CI results with respect to the number of partial waves and the
size of the one-electron basis. The results obtained for the wavelengths of the $K\alpha$
transitions improve previous theoretical predictions and compare favourably with the preliminary
results of the ongoing experiment.

\section*{Acknowledgement}

The work reported in this paper was supported by BMBF under Contract No.~05K13VHA.


\begingroup

\begin{table*}
\caption{Energy levels of beryllium-like argon Ar$^{14+}$, in Rydbergs,
1 Ry = 109\,737.315\,685\,39\,(55)~cm$^{-1}$. Separately listed are the
Dirac-Coulomb energy, the Breit correction, and the QED
correction.
The total energy is presented for the ground state, whereas
for all other states the energies relative to the ground state are given.
The theoretical contributions are presented multiplied by the reduced mass
prefactor $\mu/m$, $1-\mu/m = 0.00001373$.
\label{tab:main}
}
\begin{ruledtabular}
\begin{tabular}{llcddddddd}
\multicolumn{2}{c}{Term} &
$J$ &
\multicolumn{1}{c}{$\ \ $Coulomb} &
\multicolumn{1}{c}{$\ \ \ \ \ $Breit} &
\multicolumn{1}{c}{$\ \ \ \ \ $QED} &
\multicolumn{1}{c}{$\ \ \ \ \ $Total} &
\multicolumn{1}{c}{$\ \ \ \ \ $NIST$^a$} &
\multicolumn{1}{c}{$\ \ \ \ \ $Other theory} &
\multicolumn{1}{c}{$\ \ \ \ \ $Experiment}
\\
\hline
\\[-5pt]
%
%
$1s^22s^2$           & $^1S$     & 0   &      -758.9231 &    0.1629 &    0.1766 &      -758.5851\,(15)  \\[2pt]
$1s^22s2p$           & $^3P$     & 0   &         2.0807 &    0.0123 &   -0.0086 &         2.0844\,(9) &     2.0839 & 2.0836\,^b & 2.0837\,(16)\,^d \\
                     &           & 1   &         2.1520 &    0.0062 &   -0.0085 &         2.1497\,(12) &     2.1493 & 2.1491\,^b & 2.14931\,(9)\,^d\\
                     &           & 2   &         2.3136 &   -0.0024 &   -0.0081 &         2.3030\,(9) &     2.3026 & 2.3025\,^b & 2.3025\,(11)\,^d\\[2pt]
$1s^22s2p$           & $^1P$     & 1   &         4.1287 &    0.0018 &   -0.0083 &         4.1222\,(21) &     4.1206 & 4.1150\,^b & 4.12096\,(18)\,^d\\[2pt]
$1s2s^22p$           & $^3P^o$   & 1   &       226.3281 &   -0.1327 &   -0.0747 &       226.1206\,(42) &            &   226.151\,^c\\
$1s2s^22p$           & $^1P^o$   & 1   &       227.4630 &   -0.1492 &   -0.0745 &       227.2394\,(25) &            &   227.254\,^c \\[2pt]
$1s2s2p^2$           & $^5P$     & 1   &       226.8624 &   -0.1200 &   -0.0834 &       226.6589\,(21) &            &   226.572\,^c \\
                     &           & 2   &       226.9599 &   -0.1220 &   -0.0832 &       226.7548\,(20) &            &   226.668\,^c \\
                     &           & 3   &       227.0892 &   -0.1501 &   -0.0829 &       226.8562\,(20) &            &   226.787\,^c \\[2pt]
$1s2s2p^2$           & $^3P$     & 0   &       229.0759 &   -0.1444 &   -0.0834 &       228.8481\,(21) &            &   228.811\,^c \\
                     &           & 1   &       229.0664 &   -0.1264 &   -0.0832 &       228.8570\,(28) &            &   228.854\,^c \\
                     &           & 2   &       229.3167 &   -0.1452 &   -0.0828 &       229.0888\,(23) &            &   229.067\,^c \\[2pt]
$1s2s2p^2$           & $^3D$     & 1   &       229.1822 &   -0.1355 &   -0.0831 &       228.9634\,(30) &            &   228.971\,^c \\
                     &           & 2   &       229.0762 &   -0.1371 &   -0.0831 &       228.8560\,(35) &   229.188 &   228.880\,^c \\
                     &           & 3   &       229.1151 &   -0.1625 &   -0.0830 &       228.8696\,(35) &   228.980 &   228.917\,^c \\[2pt]
$1s2s2p^2$           & $^3S$     & 1   &       230.1192 &   -0.1319 &   -0.0831 &       229.9034\,(36) &   229.912 &   229.939\,^c \\[2pt]
$1s2s2p^2$           & $^3P$     & 0   &       230.3550 &   -0.1097 &   -0.0835 &       230.1618\,(102) &           &   230.274\,^c \\
                     &           & 1   &       230.4777 &   -0.1128 &   -0.0831 &       230.2818\,(141) &      &   230.390\,^c \\
                     &           & 2   &       230.6301 &   -0.1301 &   -0.0828 &       230.4172\,(181) &      &   230.515\,^c \\[2pt]
$1s2s2p^2$           & $^1D$     & 2   &       230.4284 &   -0.1212 &   -0.0832 &       230.2240\,(120) &   230.398 &   230.303\,^c \\[2pt]
$1s2s2p^2$           & $^1P$     & 1   &       231.3742 &   -0.1577 &   -0.0830 &       231.1334\,(161) &      &   231.256\,^c \\[2pt]
$1s2s2p^2$           & $^1S$     & 0   &       231.5411 &   -0.1120 &   -0.0829 &       231.3463\,(122) &           &   231.410\,^c \\
\end{tabular}
\end{ruledtabular}
$^a$ NIST data base \cite{nist:13},\\
$^b$ Safronova {\em et al.} \cite{safronova:96}, \\
$^c$ Costa {\em et al.} \cite{costa:01}, \\
$^d$\ Edl\'en \cite{edlen:83}.
\end{table*}
\endgroup

\begin{table}
\caption{Wavelengths of the $K\alpha$
transition lines in beryllium-like iron Ar$^{14+}$, in
  \AA.
\label{tab:trans}
}
\begin{ruledtabular}
\begin{tabular}{ld}
\multicolumn{1}{c}{Transition} &
\multicolumn{1}{c}{Wavelength}
\\
\hline
\\[-5pt]
$1s^22s2p\,^3P_{1  } \to 1s2s2p^2\,^3P_{0  }$  &   3.996\,57\, (18) \\
$1s^22s2p\,^3P_{2  } \to 1s2s2p^2\,^1D_{2  }$  &   3.998\,17\, (21) \\
$1s^22s2p\,^3P_{1  } \to 1s2s2p^2\,^3S_{1  }$  &   4.001\,11\, (10) \\
$1s^22s2p\,^3P_{2  } \to 1s2s2p^2\,^3S_{1  }$  &   4.003\,80\, (10) \\
$1s^22s^2\,^1S_{0  } \to 1s2s^22p\,^1P^o_{1  }$  &   4.010\,16\, (4)  \\
$1s^22s2p\,^1P_{1  } \to 1s2s2p^2\,^1S_{0  }$  &   4.010\,43\, (22) \\
$1s^22s2p\,^1P_{1  } \to 1s2s2p^2\,^1P_{1  }$  &   4.014\,19\, (29) \\
$1s^22s2p\,^3P_{1  } \to 1s2s2p^2\,^3D_{1  }$  &   4.017\,69\, (6)  \\
$1s^22s2p\,^3P_{2  } \to 1s2s2p^2\,^3P_{2  }$  &   4.018\,18\, (5)  \\
$1s^22s2p\,^3P_{0  } \to 1s2s2p^2\,^3P_{1  }$  &   4.018\,42\, (5)  \\
$1s^22s2p\,^3P_{1  } \to 1s2s2p^2\,^3D_{2  }$  &   4.019\,59\, (7)  \\
$1s^22s2p\,^3P_{2  } \to 1s2s2p^2\,^3D_{1  }$  &   4.020\,41\, (6)  \\
$1s^22s2p\,^3P_{2  } \to 1s2s2p^2\,^3D_{3  }$  &   4.022\,07\, (7)  \\
$1s^22s2p\,^3P_{2  } \to 1s2s2p^2\,^3P_{1  }$  &   4.022\,29\, (5)  \\
$1s^22s2p\,^3P_{2  } \to 1s2s2p^2\,^3D_{2  }$  &   4.022\,31\, (7)  \\
$1s^22s2p\,^1P_{1  } \to 1s2s2p^2\,^3P_{2  }$  &   4.026\,90\, (32) \\
$1s^22s^2\,^1S_{0  } \to 1s2s^22p\,^3P^o_{1  }$  &   4.030\,00\, (8)  \\
$1s^22s2p\,^1P_{1  } \to 1s2s2p^2\,^1D_{2  }$  &   4.030\,34\, (22) \\
$1s^22s2p\,^3P_{2  } \to 1s2s2p^2\,^5P_{3  }$  &   4.058\,13\, (4)  \\
\end{tabular}
\end{ruledtabular}
\end{table}

\end{document}